\documentclass[useAMS,usenatbib]{mn2e}

\usepackage{graphicx}
\usepackage{natbib}
\usepackage{aas_macros}

\begin{document}
\title[The MS of SFGs at z$\sim$0.6: reinstating major mergers]{The main sequence of star-forming galaxies at z$\sim$0.6: reinstating major mergers}
\author[M. Puech et al.]{M. Puech\thanks{E-mail:
mathieu.puech@obspm.fr}, F. Hammer, M. Rodrigues, S. Fouquet, H. Flores, K. Disseau\\
GEPI, Observatoire de Paris, CNRS, University Paris Diderot; 5 Place Jules Janssen, 92190 Meudon, France}

\date{Accepted ... Received ...}

\pagerange{\pageref{firstpage}--\pageref{lastpage}} \pubyear{2002}

\maketitle

\label{firstpage}

\begin{abstract}
  The relation between the star formation rate and the stellar mass of
  star-forming galaxies has been used to argue that major mergers
  cannot be the main driver of star formation. Here, we re-examine
  these arguments using the representative IMAGES-CDFS sample of
  star-forming galaxies at $z=0.4-0.75$, taking advantage of their
  previously established classification into pre-fusion, fusion, and
  relaxing galaxy mergers. Contrary to previous claims, we show there
  is no tension between the main sequence scatter and the
  \emph{average} duration of the fusion star formation rate $SFR$
  peak. We confirm previous estimates of the fraction of $SFR$ due to
  morphologically-selected galaxies ($\sim$23\%) or the $SFR$
  enhancement due to major merger during the fusion phase
  ($\sim$10\%). However, galaxy mergers are not instantaneous
  processes, which implies that the total fraction of the $SFR$
  associated to galaxies undergoing major mergers must account for the
  three merger phases. When doing so, galaxies involved in major
  mergers are found to represent 53-88\% of the total $SFR$ at
  $z\sim0.6$. The fraction of LIRGs in the fusion phase is found to be
  in agreement with the observed morphological fraction of LIRGs
  without disks and with the observed and expected major merger rates
  at $z \leq 1.5$.
\end{abstract}


\section{Introduction}
The so-called ``SFR-sequence'' or ``Main Sequence'' (hereafter, MS) of
star forming galaxies (SFGs) is a relatively tight correlation between
the star formation rate $SFR$ and stellar mass $M_{stellar}$. The MS
appears to hold at least up to $z\sim2.5$, with a strong evolution in
zero point (e.g., \citealt{noeske07,elbaz07,rodighiero10}), and a
possible flattening above \citep{rodighiero11,whitaker12}. The exact
shape of the MS remains uncertain at the highest masses
\citep{brinchmann04,drory08,whitaker12}, as well as at low $SFR$ due
to the presence of a ``cloud'' of low $SFR$ galaxies extending
downward the MS \citep{wuyts11,fumagalli13}. Nevertheless, the MS has
been used as a typical region of the $SFR$-$M_{stellar}$ plane within
which analytic, semi-analytic, or cosmological models are expected to
produce the most typical SFGs at a given redshift $z$ (e.g.,
\citealt{bouche10,dutton10,kannan13}).

A lot of interest has been devoted to the scatter of the MS, which is
found to be $\sim$ 0.3 dex independent of $z$ (e.g.,
\citealt{rodighiero11,whitaker12}), though with a possible evolution
in mass \citep{guo13}. This relatively small scatter has been used to
estimate the average duty cycle of episodic star formation episodes:
$SFR$ variations exceeding $\pm$1(2)$\sigma$=0.3(0.6) dex (i.e.,
factors 2(4)) should have duty cycles $<$32(5)\%, meaning that on
average galaxies with $M_{stellar}\sim 10^{10-11}M_\odot$ could not
have spent more than $\sim$2.5(0.4)Gyr in episodes of enhanced star
formation since $z=1$. $SFR$ enhancements in simulated major mergers
are found to be too large in amplitude and too short in duration to
statistically account for the MS scatter (e.g.,
\citealt{cox08,lotz10}). This was used to argue that major mergers
cannot account for the MS scatter. This argument was further supported
by the enhancement of the star formation activity directly measured in
morphologically-selected major mergers, with typically less than
$\sim$ 10\% of the $SFR$ density directly triggered by major mergers
\citep{robaina09,rodighiero11}, and by the relatively large fraction
($\sim$50\%) of Luminous InfraRed Galaxies (hereafter LIRGs) found to
harbor disk morphologies at $z\leq1$ (e.g.,
\citealt{zheng04,melbourne05}).

Other observations, in particular from the 3D survey IMAGES, have
suggested that gas-rich major mergers played a prominent role in the
structural evolution of intermediate mass galaxies since $z\sim 1$
\citep{hammer05,hammer09,hopkins09}. The IMAGES sample is therefore
ideal to re-examine these issues and test whether the MS scatter-based
duty cycle argument and the insights gained on the structural
evolution of intermediate-mass galaxies at $z\leq1$ using 3D surveys
are in tension over the past 9 Gyr. Throughput this letter, we used a
``concordance'' cosmological model when needed with ($\Omega _m$,
$\Omega _\Lambda$, $h$)=(0.3, 0.7, 0.7), $AB$ magnitudes, and a diet
Salpeter IMF \citep{bell03}.

\section{The average $\MakeLowercase{z}\sim0.6$ major merger and the MS}

We started from the IMAGES sub-sample of 35 star-forming galaxies that
lie in the CDFS region, which is representative of the SFG population
with $M_J \leq -20.3$ and $EW_0([OII])\geq15\AA$ at $z\sim0.6$
\citep{yang08}. Based on their observed spatially-resolved morphology
(using the ACS camera on-board HST) and kinematics (using the
multi-IFU VLT/FLAMES-GIRAFFE optical spectrograph), this sample was
classified into three distinct morpho-kinematic classes, namely
Rotators (ROT), which correspond to galaxies that were classified as
spiral and rotating disks according to their morphology and kinematics
\citep{neichel08}, Non-Relaxed (NR) systems, for which the morphology
and the kinematics was found to be peculiar, and Semi-Relaxed (SR)
systems, which correspond to galaxies that do not meet the two
previous criteria (i.e., with a relaxed morphology and non-relaxed
kinematics or \emph{vice-versa}; see \citealt{hammer09}). Two galaxies
were rejected because they turned out to be outliers, while another
one was rejected because it did not have HST imaging (see
\citealt{hammer09} and \citealt{puech12} for details), which led to
the sample of 32 galaxies studied here. Stellar masses and $SFR$s were
estimated in \cite{puech08,puech10} using simplified prescription
between J-band mass-to-light ratios and color from \cite{bell03},
while $SFR$ were estimated by summing the UV-unobscured and IR-based
contributions (with an average uncertainty of $\sim$33\%). LIRGs were
identified as galaxies with counterparts within 1 arcsec of the 24$\mu
m$ MIPS DR3 public catalog and $SFR\geq12.2M_\odot/yr$
\citep{lefloch05}.

\begin{figure}
\centering
\includegraphics[width=9cm]{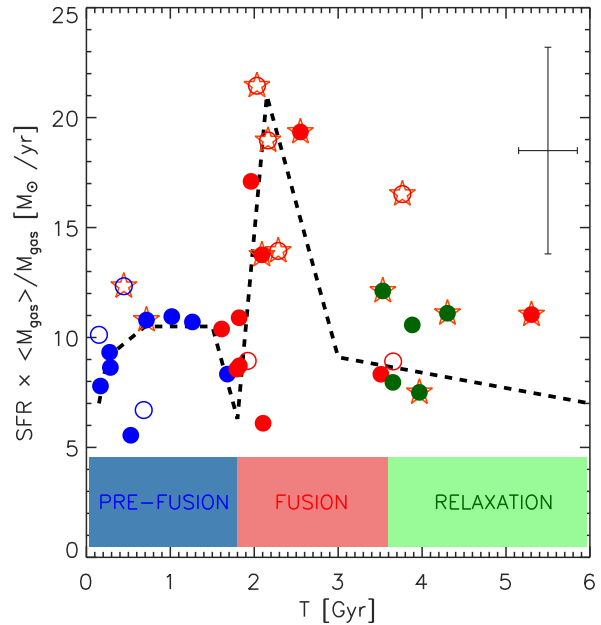}
\caption{Normalized $SFR$ as a function of time for the IMAGES-CDFS
  galaxies. Each galaxy is plotted at the time at which individual
  numerical models give the best fit to observations. The $SFR$ is
  normalized by the estimated gas mass (see \citealt{puech12}) to
  remove biases due to the different gas reservoir contents from
  object to object. Full symbols represent galaxies whose
  morpho-kinematics compared best with those of their models (secured
  cases, see text). Merging galaxies were classified into three
  different classes: the pre-fusion (blue symbols), post-fusion (red),
  and relaxation phases (green). The dash line is not a fit but a
  simple visual guide through the points. The median uncertainty is
  indicated in the upper-right corner. The errorbar on the merging
  timescale corresponds to the time-step between two simulated
  snapshots in the models. LIRGs are indicated by orange open stars.}
\label{figseq}
\end{figure}

The observed spatially-resolved morpho-kinematic properties of the 27
NR and SR galaxies in the IMAGES-CDFS sample were modeled using a grid
of hydrodynamical gas-rich major mergers. The best model was graded by
three examiners as described in \cite{hammer09}, and secure major
merger candidates were identified as galaxies having a model that
fitted observations with a good level of confidence (i.e., grades
$\geq$ 4/6), which represent 18 cases out of 27. The remaining 9
objects are either major mergers that could not be secured with a high
enough grade because of the limited size of the simulation grid, or
galaxies possibly undergoing other evolutionary processes such as
minor mergers (e.g., \citealt{puech07}) or internal instabilities
(e.g., \citealt{puech10b}). \cite{puech12} showed that the resulting
major merger rate (for baryonic mass ratios larger than 0.25) is found
to be in remarkable agreement with predictions from semi-empirical
$\Lambda$-CDM models. The key factor for this agreement is that the
combination of morphology with spatially-resolved kinematics is found
to be sensitive to all the phases of the merging process (see Fig.
\ref{figseq}), from the pre-fusion phase during which the two
progenitors can still be identified as distinct components, the
fusion/post-fusion phase during which they coalesce and generally
result in a peak of $SFR$, to the relaxation phase in which the
remnant progressively reaches a relaxed dynamical state. Since the
IMAGES-CDFS is representative of the galaxy population at $z\sim0.6$
and that all the merger phases are well-sampled, Fig. \ref{figseq}
illustrates that this sample as a whole can be used to represent the
average $z\sim0.6$ major merger.

The average pre-fusion and fusion phases at $z\sim0.6$ were found to
last 1.8 Gyr, in agreement with expectations from hydrodynamical
simulations (e.g., \citealt{cox08}). This is consistent with the
average $SFR$ enhancement duration inferred from the MS scatter, which
is estimated to be 0.4-2.5 Gyr (from the 1-2$\sigma$ scatter, see
Sect. 1). Indeed, all on-going major mergers at $z=0.4-0.75$ span a
range of separations, mass ratios, orbits, and gas fractions, which
result in a range of different individual pre-fusion and fusion phases
durations \citep{lotz10,lotz10b}. Statistically, the average fusion
phase duration corresponds to the envelope of the individual fusion
peaks with a resulting width much larger than the width of the
individual fusion peaks. Figure \ref{figseq} shows the resulting
average fusion peak in the IMAGES-CDFS sample, with an $SFR$
enhancement in amplitude of a factor $\sim$2-3. Both the
\emph{average} $SFR$ enhancement amplitude and duration are therefore
in very good agreement with constraints inferred from the MS scatter
(see Sect. 1). There is therefore no tension between the $SFR$
constraints inferred from the scatter of the MS and major mergers,
\emph{provided that the proper empirical statistics on separations,
  mass ratios, orbits, and gas fractions is considered}.

\begin{figure}
\centering \includegraphics[width=7.cm]{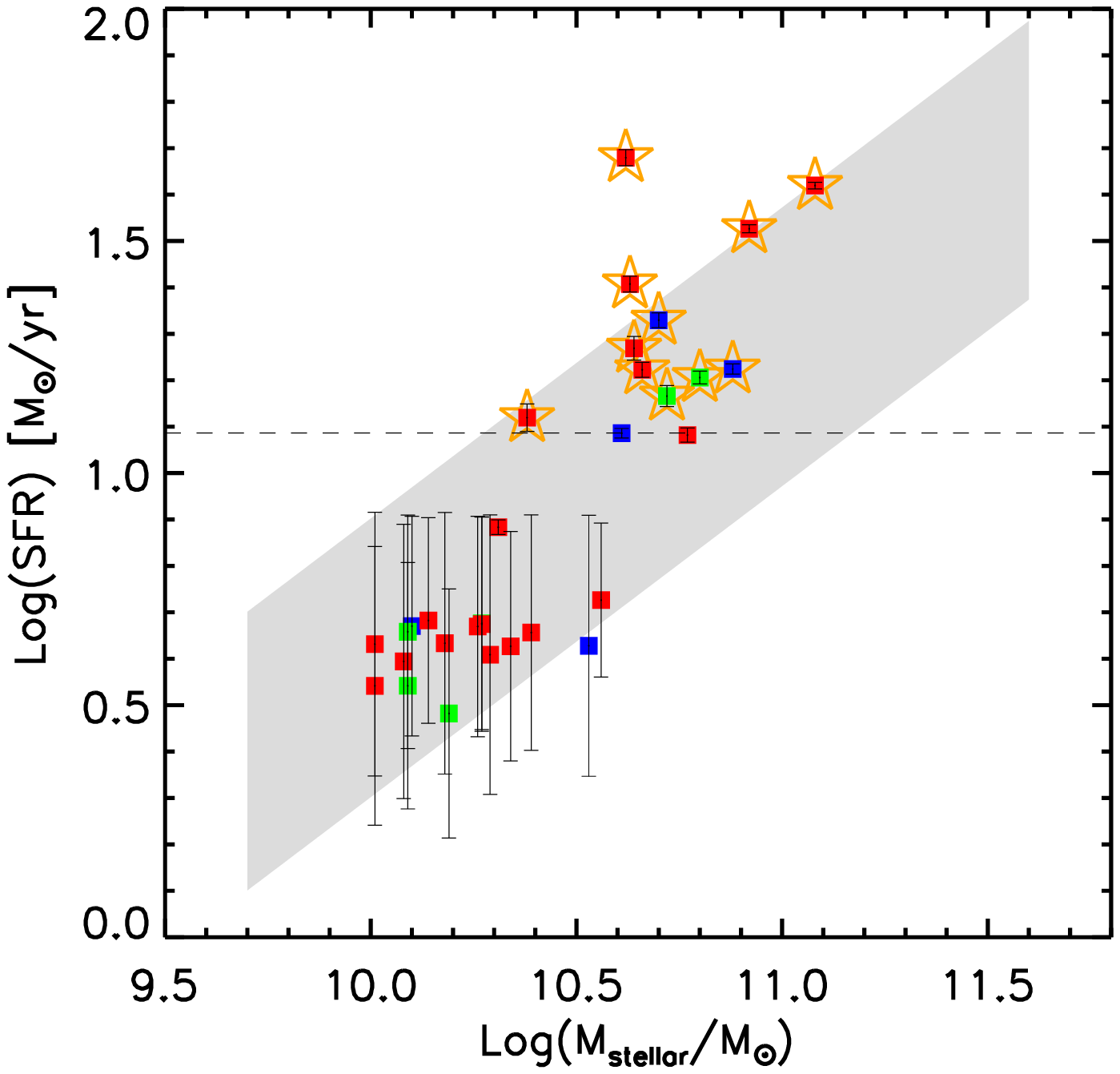}
\centering \includegraphics[width=7.cm]{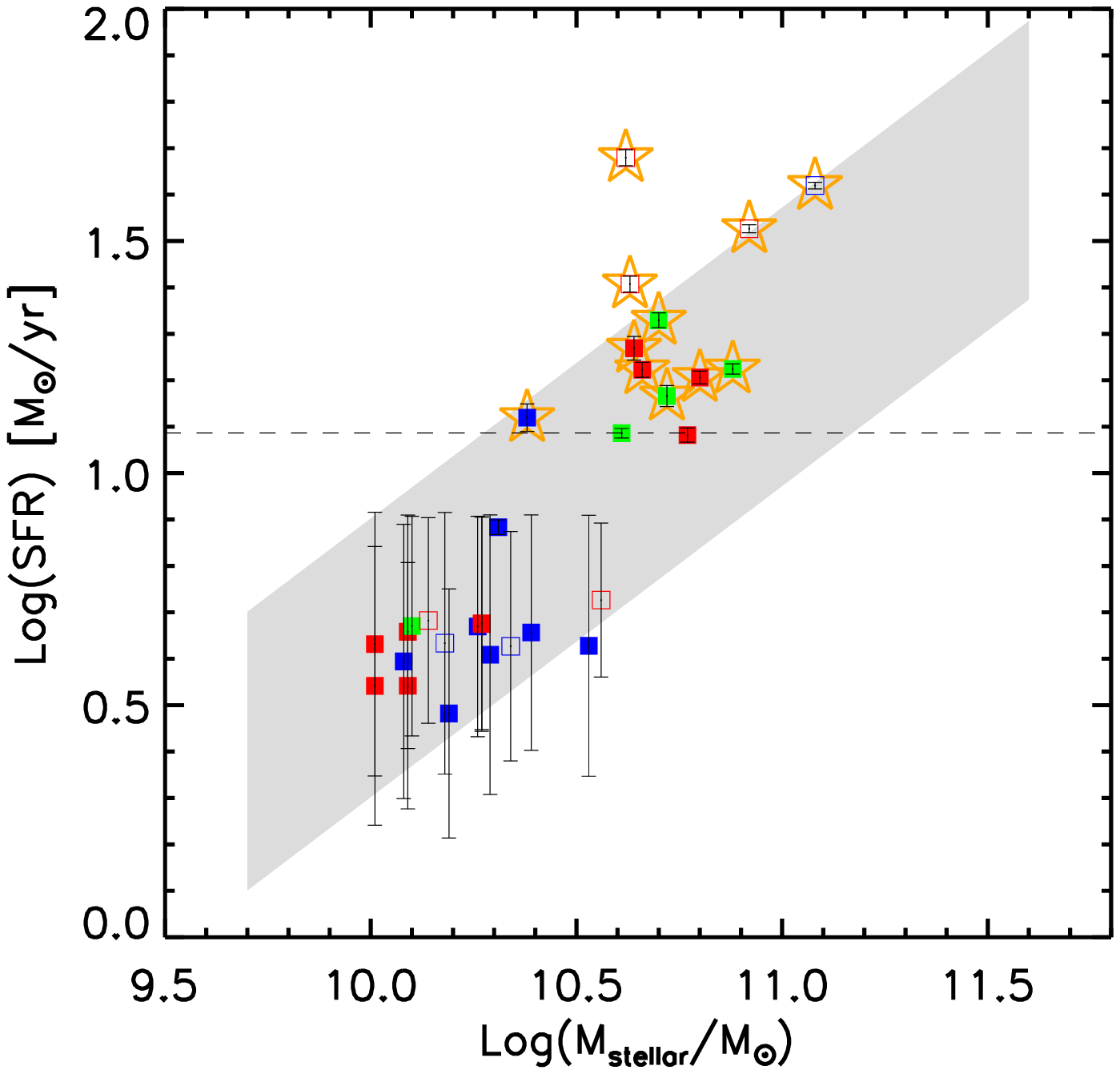}
\caption{IMAGES-CDFS MS as a function of the morpho-kinematic
  classification (\emph{top panel}) and merger phase (\emph{bottom
    panel}). The grey region represents the $z=0.45-0.7$ MS and its
  $\pm$1-$\sigma$ scatter from Noeske et al. (2007), as reported by
  Dutton et al. (2010). \emph{Top panel:} Red squares represent NR
  galaxies, green squares SR galaxies, and blue squares rotators.
  \emph{Bottom panel:} Blue squares represent galaxies in the
  pre-fusion phase, red squares galaxies in the fusion phase, while
  green squares represent galaxies in the relaxation phase (see Fig.
  \ref{figseq}). Full symbols represent galaxies with secured models
  (see text). LIRGs are indicated as orange open stars (with errorbars
  typically smaller than the symbol size), and the corresponding SFR
  threshold indicated as a black dash line.}
\label{msclass}
\end{figure}

Figure \ref{msclass} shows how galaxies are distributed across the MS
as a function of their morpho-kinematic classification and merger
phase. It reveals that NR galaxies tend to lie on average above the
MS, while SR and ROT lie closest to the mean relation, which is
consistent with other observations using morphology as a tracer of the
galaxy dynamical state \citep{jogee09,kaviraj13,hung13}, or
spatially-resolved kinematics \citep{green13}. Interestingly, galaxies
lying above the 1-$\sigma$ scatter are all galaxies with NR
morpho-kinematics in the fusion/post-fusion phase, as predicted by the
semi-empirical model of \cite{hopkins10}. In this model, such galaxies
are expected to be mostly in the fusion peak, which, on average,
enhance their SFRs. Conversely, galaxies populating the MS are a mix
of the three merger phases. Previous studies claimed that the MS is
not related to major mergers because most of morphologically-selected
mergers lie above the relation. \emph{This is true only if ones limits
  major mergers to the fusion phase}. On the contrary, we find that
the MS is mostly populated by galaxies with secured major merger
models, though corresponding mostly to the pre-fusion and relaxation
phases, i.e., to the less star forming phases.

\section{Which fraction of the $SFR$ is really related to major mergers?}

We estimated the fraction of $SFR$ associated to each merger phase.
First, we considered galaxies with secured major merger models only.
Because part of the galaxies with unsecured models could also be due
to major mergers that were not identified by the limited grid of
models (see Sect. 2), this sample provides a strict lower limit to the
fraction of $SFR$ associated to major mergers (see first column of
Tab. \ref{tabfrac}). To partly account for possible major mergers
missed by the grid of models, we also estimated these fractions
considering secured models with addition of the five galaxies with
unsecured models in the fusion phase. Galaxies in the relaxation phase
are indeed the most difficult to identify from morpho-kinematic data
because most of them lack of strong disturbances. In the case of these
five galaxies, their best-models were systematically found close to
the end of the simulation \citep{puech12}, which might indicate that
they actually belong to the relaxation phase but were misclassified
because of the inherent difficulty of identifying galaxies in this
phase and/or the limited grid of simulations used for comparison with
observations. The corresponding fractions are listed in the second
column of Tab. \ref{tabfrac}. The third column gives a median between
the two estimates. Since the IMAGES-CDFS sample is representative of
the $z\sim 0.6$ emission line galaxy population (see Sect. 2), this
provides lower limits to the average fraction of $SFR$ associated to
each major merger phase at $z\sim 0.6$.

\begin{table}
\centering
\begin{tabular}{cccc}\hline
Merger Phase & & \% of $SFR$ & \\
      & Secured only & Extended & Median\\\hline
Pre-fusion & 12$\pm$2 & 12$\pm$2 &  12$\pm$2\\ 
Fusion & 23$\pm$5 & 23$\pm$5 & 23$\pm$5\\
Relaxation & 53$\pm$11 & 18$\pm$3 & 35$\pm$17\\\hline
Unsecured & 12$\pm$5 & 47$\pm$13 & 30$\pm$17\\\hline
Enhancement & 8$\pm$5 & 11$\pm$5 & 10$\pm$5\\\hline
\end{tabular}
\caption{Fraction of $SFR$ in the IMAGES-CDFS sample as a function of the merger phase (pre-fusion, fusion, and relaxation phase); the fourth row represents the contribution from galaxies with unsecured major merger models. Uncertainties were estimated using bootstrap re-sampling. In the first column, only galaxies with secured models were considered, while the second column correspond to secured models with addition of five galaxies in the relaxation phase (see text). The right column is a median between the first two. The last line gives the fraction of $SFR$ associated to the fusion peak enhancement.}
\label{tabfrac}
\end{table}

We find that the fraction of $SFR$ in each merger phase is generally
consistent with results from the literature (e.g.,
\citealt{robaina09}). The fraction of $SFR$ associated to all galaxies
involved in major mergers (i.e., regardless of the merger phase) is
found to be 53-88\% (70\% in median; see Tab. \ref{tabfrac}). That is
not to say that such a large fraction of the $SFR$ is triggered by
major mergers \emph{only}. Indeed, the $SFR$ in each merger phase can
be triggered by processes other than the merger itself, such as
internal instabilities, minor mergers, or gas accretion. This
degeneracy is minimized during the fusion peak, in which the $SFR$ is
mostly driven by the merger itself since it is during this phase that
the tidal torques drive a fraction of the gas inward and results in a
central starburst \citep{hopkins09}. Galaxies in the fusion phase are
found to be responsible for 23$\pm$5\% of the total $SFR$.
Morphogically-selected mergers (i.e., mostly galaxies in the fusion
peak; \citealt{lotz10,hopkins10}), are consistently found to account
for 15-21\% of the $SFR$ (\citealt{bell05}, \citealt{robaina09}).
\cite{jogee09} also reported that morphologically-selected major
mergers contribute $<$30\% of the total $SFR$ at similar redshifts and
masses.

In order to roughly correct this fraction from contributions of other
processes, one can tentatively subtract to the fraction of $SFR$ in
the fusion phase the average contribution from galaxies in the two
other phases to estimate the $SFR$ peak enhancement due to major
mergers only. This leads to a fraction of 10$\pm$5\% (see Tab.
\ref{tabfrac}), which is consistent with results reported by
\cite{robaina09}, who find that 8$\pm$3\% of the $SFR$ is directly
triggered by major merger in a similar range of mass and redshift (see
also \citealt{jogee09}. This is again a lower limit of the $SFR$
fraction due to major mergers only since part of the pre-fusion and
relaxation phase $SFR$ is also triggered by the merger event itself:
the first passage between the two progenitors can also result in a
small peak of $SFR$ as suggested by hydrodynamical simulations
\citep{cox08,lotz10}, while part of gas expelled during the merger can
fall back onto the remnant and reform a stellar disk (e.g.,
\citealt{robertson06,hopkins09}).

The total $SFR$ fraction associated to major mergers \emph{cannot} be
derived by selecting galaxies in the fusion peak only, i.e., by
considering only the most morphologically disturbed galaxies, or even
by considering the $SFR$ enhancement during the fusion peak: this
would be equivalent to assume that galaxies merge and relax almost
instantaneously, which is clearly at odd with both simulations (e.g.,
\citealt{cox08,lotz10,lotz10b}) and observations (e.g.,
\citealt{rothberg04}; see also Fig. 1). From a causal point of view, one
cannot avoid to consider what happens \emph{before} and \emph{after}
the coalescence of the two progenitors. One must consider \emph{all}
the merger phases (see Fig. \ref{figseq}) to derive the proper
fraction of $SFR$ related to mergers.

\section{Discussion: The fraction and morphological split of LIRGs}
As expected, LIRGs correspond to the most SFGs in the sample (see Fig.
\ref{msclass}), and represent 73$\pm$5\% of the total $SFR$
\citep{flores99,lefloch05}. This strengthens that the [OII] selection
in the IMAGES sample results in a representive sample in terms of
total $SFR$, and that no significant obscure sources were missed (see
Sect. 2). A large fraction of LIRGs was found to be associated
to disky morphologies, which has been used to claim that star
formation in LIRGs is not mainly driven by major mergers (e.g.,
\citealt{bell05, melbourne05}). We find that 58$^{+17}_{-10}$\% of
LIRGs in the IMAGES-CDFS sample are in the fusion phase. Since most
morphologically-selected mergers are found to correspond to the fusion
phase, it implies that the remaining $\sim$42\% of LIRGs that lie in
the pre-fusion and relaxation phase should correspond to less
disturbed morphology and kinematics. This matches very well the
observed fraction of LIRGs that is classified as disks at the same
redshift and mass range, with, e.g., $\sim$42-47\% as reported by,
e.g., \cite{zheng04} and \cite{melbourne05}. This is consistent with
the position of LIRGs across the MS, since the disky LIRGs are
preferentially found along the MS, while those in interaction
preferentially lie above the MS \citep{hung13}; Figure \ref{msclass}
reveals the same trend in the IMAGES-CDFS LIRGs, although with much
smaller statistics. We note that compact LIRGs are preferentially
found to lie above the MS \citep[at least at $z\sim$0, see Fig. 16
of][]{elbaz11}, which is consistent with these LIRGs being in the
fusion peak, where most of the material falls towards the center of
mass of the merging system \citep{hammer05}.

If one assumes that all intermediate-mass galaxies experienced an
infrared episode (IRE) due to intense star formation activity, then
the average number of such episodes per galaxy is $n_{IRE}=0.15\times
0.58^{+0.17}_{-0.10}\times \Delta t / \tau_{IRE}$, where $\Delta t$ is
the ellapsed time over a given redshift range, 15\% if the fraction of
LIRGs in the same range of mass and redshift, and $\tau_{IRE}$ is the
average timescale of the IRE (see \citealt{hammer05}). With
$\tau_{IRE}=\tau_{fusion}=1.8\pm0.1$ Gyr and over $\Delta z$=0-1
(resp., $\Delta z$=0-1.5), this yields $n_{IRE}=37^{+13}_{-8}$\%
(resp., $45^{+16}_{-11}$\%), which compares very well with the
expected and observed fraction of local galaxies having experienced a
major merger over the same redshift
range, i.e., 35\% (resp., 50\%; \citealt{puech12}).\\

Both theoretically and observationally, gas-rich major mergers were
found to play a dominant role in the structural evolution of
present-day spiral galaxies over the past 9 Gyr (e.g.,
\citealt{hopkins10,puech12}). Here we have shown that they also
account for a major fraction of the $SFR$ in their ancestors and drive
the MS of SFGs. This major merger-driven star formation activity is
consistent with the observed fraction and morphology of LIRGs, in
agreement with the ``spiral rebuilding scenario'' initially proposed
by \cite{hammer05}.

\section*{Acknowledgments}
We thank Susanna Vergani and Hakim Atek for useful discussions and
comments. We thank the anonymous referee for a careful reading and
valuable comments, which have significantly contributed to improve the
clarity of the paper.

%\begin{thebibliography}{99}
%\bibitem[\protect\citeauthoryear{Baird}{1981}]{b1} Baird S.R., 1981,
%ApJ, 245, 208
%\end{thebibliography}
%%%%%%%%%%%%%%%%%%%%%%%%%%%%%%%%%%%%%%%%%%%%%%%%%%%%%%%%%%%%%
%%%%% References %%%%%
\bibliographystyle{mn2e}
\bibliography{mainseq_astroph}

%\bsp

\label{lastpage}

\end{document}